\documentclass[12pt]{iopart}

\usepackage{graphics}

\usepackage{amssymb}

\begin{document}

\title{Time evolution of a superposition of dressed oscillator
states in a cavity}

\author{G Flores-Hidalgo$^1$, C A Linhares$^2$, A P C Malbouisson$^3$
and J M C Malbouisson$^4$}

\address{$^1$ Instituto de Ci\^encias Exatas, Universidade Federal de
Itajub\'a, 37500-903, Itajub\'a, MG, Brazil}
\address{$^2$ Instituto de F\'{\i}sica, Universidade do Estado do Rio
de Janeiro, 20559-900, Rio de Janeiro, RJ, Brazil}
\address{$^3$ Centro Brasileiro de Pesquisas F\'{\i}sicas/MCT,
22290-180, Rio de Janeiro, RJ, Brazil}
\address{$^4$ Instituto de F\'{\i}sica, Universidade Federal da
Bahia, 40210-340, Salvador, BA, Brazil}

\eads{\mailto{gfloreshidalgo@unifei.edu.br},
\mailto{linharescesar@gmail.com},
\mailto{adolfo@cbpf.br},\mailto{jmalboui@ufba.br}}

\begin{abstract}
Using the formalism of {\it renormalized} coordinates and
\textit{dressed} states introduced in previous publications, we
perform a nonperturbative study of the time evolution of a
superposition of two states, the ground state and the first excited
level of a harmonic oscillator, the system being confined in a
perfectly reflecting cavity of radius $R$. For $R\rightarrow\infty$,
we find dissipation with dominance of the interference terms of the
density matrix, in both weak- and strong-coupling regimes. For small
values of $R$ all elements of the density matrix present an
oscillatory behavior as times goes on and the system is not
dissipative. In both cases, we obtain improved theoretical results
with respect to those coming from perturbation theory.
\end{abstract}


\submitto{\JPA}

\maketitle

\section{Introduction}

The main analytical method used to treat the physics of interacting
systems is perturbation theory. In this framework, {\it bare},
noninteracting,  matter fields and gauge fields (to which bare
quanta are associated) are introduced, the interaction being
accounted for order by order in powers of the coupling constant in
the perturbative expansion of the observables. For a long time,
physicists have been aware that perturbative expansions have a
limited scope, in spite of all the remarkable achievements obtained
with them. Some examples of such situations are resonant effects
associated to the coupling of atoms to strong radiofrequency fields,
or the low-energy domain of quantum chromodynamics, where
confinement of quarks and gluons takes place. Along the last
decades, many attempts were devised to circumvent the limitations of
perturbation theory, in particular when strong effective couplings
are involved.

In any case, as a matter of principle, due to the nonvanishing of
the coupling constant, the idea of a bare particle associated to a
bare matter field is actually an artifact of perturbation theory and
strictly speaking is physically meaningless. In general, a charged
physical particle is always coupled to a gauge field; in other
words, it is always ``dressed'' by a cloud of quanta of the gauge
field, for instance, photons in the case of electrodynamics.

With respect to the time evolution of systems of (matter) particles,
the idea is that particles are coupled to an environment or to a
thermal bath. There are usually two equivalent ways of modeling the
environment to which the particle is coupled: either to represent it
by a free field, as was done in Refs.~\cite{zurek,paz}, or to
consider the environment as a reservoir composed of a large number
of noninteracting harmonic oscillators (see, for instance,
\cite{ullersma,haake,caldeira,schramm}). In both cases, exactly the
same type of argument given above for a charged particle applies,
making the appropriate changes, to these systems. We may speak of a
``dressing'' of the set of particles by the ensemble of the harmonic
modes of the environment. The particles in the system are considered
as always ``dressed'' by a cloud of quanta of the environment. This
is true in general for any system in which material particles are
coupled to an environment, no matter what the specific nature of the
environment and interaction involved are.

In the present work we adopt this point of view and consider a
harmonic oscillator, coupled linearly to an environment modeled by
the infinite set of harmonic modes of a scalar field, the whole
system being contained in a perfectly reflecting sphere of radius
$R$. We implement a nonperturbative study of the time evolution of a
superposition of the first excited and the ground states of the
oscillator, by means of the \emph{dressed} states introduced in
\cite{adolfo1} and already employed to investigate several
situations \cite{adolfo2,adolfo3,adolfo4,adolfo5,ga1,ga2}. Such a
nonperturbative approach is possible due to the linear character of
the interaction between the oscillator and the environment we
choose.

The semiqualitative idea of a ``dressed atom" was originally
introduced in Refs.~\cite{Polonsky} and \cite{Haroche}, and largely
employed in studies involving the interaction of atoms and
electromagnetic fields \cite{Cohen1,Cohen2,Haroche1,Haroche2}. In
the realm of general physics, the concept of dressing a matter
particle by an environment has found an application  in describing
the radiation damping of classical systems \cite{petrosky}. Our
dressed states can be viewed as a rigorous version of these dressing
procedures. Moreover, the oscillator may represent a mode of the
quantized electromagnetic field in a cavity interacting with the
environment; in this case, our study refers to the time evolution of
the superposition of the ground state and the first excited state of
the field in the cavity.

The model in itself is not new; a system composed of a material body
interacting linearly with an environment has been the main subject
of many papers, as those quoted in Refs.
\cite{zurek,paz,ullersma,haake,caldeira,schramm,ford,petrosky}. The
novelty lies in the nonperturbative approach to the problem, by
means of {\it renormalized} coordinates and {\it dressed} states,
which was started in \cite{adolfo1}. In particular, our dressed
states are {\it not} the same as those employed in the literature,
usually associated to normal coordinates. Our dressed states are
given in terms of our  {\it renormalized} coordinates and allow a
rigorous study of  the time evolution of quantum systems. The
results we obtain by this means are those expected on physical
grounds, but contain corrections with respect to the formulas
obtained from perturbation theory.

We structure the paper as follows. In Section II we review the basic
aspects of the formalism of renormalized coordinates and dressed
states and then, in Section III, we treat the time evolution of a
superposition of oscillator states in both limiting situations of
$R\rightarrow\infty$ and small $R$. Section IV contains our final
comments.

\section{Renormalized coordinates and dressed states}

Let us start by considering the harmonic oscillator having
\emph{bare} frequency $\omega _0$, linearly coupled to a field
described by $N$ ($\rightarrow\infty$) other oscillators, with
frequencies $\omega _k$, $k=1,2,\ldots ,N$. The whole system is
contained in a perfectly reflecting spherical cavity of radius $R$,
the free space corresponding to the limit $R\rightarrow \infty $.
Hereafter, we shall refer to the harmonic oscillator as the {\it
particle}, to distinguish it from the harmonic modes of the
environment. Denoting by $q_0(t)$ ($p_0(t)$) and $q_k(t)$ ($p_k(t)$)
the coordinates (momenta) associated with the particle and the field
oscillators respectively, the Hamiltonian of the system is
\begin{equation}
H=\frac 12\left[ p_0^2+\omega _0^2q_0^2+\sum_{k=1}^N\left(
p_k^2+\omega _k^2q_k^2\right) \right] - q_0\sum_{k=1}^N \eta
\omega_k q_k , \label{Hamiltoniana}
\end{equation}
where the limit $N\rightarrow \infty $ is understood and $\eta$ is a
constant.

The Hamiltonian (\ref{Hamiltoniana}) can be turned to principal axis
by means of a point transformation,
\begin{equation}
q_\mu  = \sum_{r=0}^{N} t_\mu ^r Q_r\,\,,\,\,\,\,\, p_\mu = \sum_{r=0}^{N} t_\mu ^r P_r \;,
\label{transf}
\end{equation}
where $\mu =(0,\{k\}),\, k=1,2,...,N$, performed by
an orthonormal matrix $T=(t_\mu ^r)$. The subscripts $\mu =0$ and
$\mu =k$ refer respectively to the particle and the harmonic modes
of the field and $r$ refers to the normal modes. In terms of normal
momenta and coordinates, the transformed Hamiltonian reads
\begin{equation}
H=\frac 12\sum_{r=0}^N(P_r^2+\Omega _r^2Q_r^2),  \label{diagonal}
\end{equation}
where the $\Omega _r$'s are the normal frequencies corresponding to
the collective \textit{stable} oscillation modes of the coupled
system.

Using the coordinate transformation (\ref{transf}) in the
equations of motion and explicitly making use of the normalization
condition
\begin{equation}
\sum_{\mu =0}^N(t_\mu ^r)^2=1 , \label{normcond}
\end{equation}
we get
\begin{equation}
t_k^r=\frac{\eta \omega_k}{\omega _k^2-\Omega
_r^2}t_0^r\;,\;\;t_0^r=\left[ 1+\sum_{k=1}^N\frac{\eta^2
\omega_k^2}{(\omega _k^2-\Omega _r^2)^2}\right] ^{-\frac 12},
\label{tkrg1}
\end{equation}
with the condition
\begin{equation}
\omega _0^2-\Omega _r^2=\sum_{k=1}^N\frac{\eta^2 \omega_k^2}{\omega
_k^2-\Omega _r^2}. \label{Nelson1}
\end{equation}
The right hand side of Eq.~(\ref{Nelson1}) diverges in the limit
$N\rightarrow\infty$. Defining the counterterm $\delta \omega^2 = N
\eta^2$, it can be rewritten in the form
\begin{equation}
\omega _0^2-\delta \omega ^2-\Omega _r^2=\eta ^2\Omega
_r^{2}\sum_{k=1}^N\frac 1{\omega _k^2-\Omega _r^2} .
\label{Nelson2}
\end{equation}

Equation (\ref{Nelson2}) has $N+1$ solutions, corresponding to the
$N+1$ normal collective modes. It can be shown \cite{adolfo1} that
if $\omega _0^2>\delta \omega ^2$, all possible solutions for
$\Omega ^2$ are positive, physically meaning that the system
oscillates harmonically in all its modes. On the other hand, when
$\omega _0^2<\delta \omega ^2$, one of the solutions is negative and
so no stationary configuration is allowed. Nevertheless, in a
different context, it is precisely this runaway solution that is
related to the existence of a bound state in the Lee--Friedrichs
model. This solution is considered in Ref.~ \cite{Likhoded} in the
framework of a model to describe qualitatively the existence of
bound states in particle physics.

Therefore, we just consider the situation in which all normal modes
are harmonic, which corresponds to the first case above, $\omega
_0^2>\delta \omega ^2$, and define the \textit{renormalized}
frequency
\begin{equation}
\bar{\omega}^2 = \lim_{N\rightarrow\infty} \left( \omega _0^2 - N
\eta^2 \right) , \label{omegabarra}
\end{equation}
following the pioneering work of Ref.~\cite{Thirring}. In the limit
$N\rightarrow \infty $, Eq.~(\ref{Nelson2}) becomes
\begin{equation}
\bar{\omega}^2-\Omega ^2=\eta ^2\sum_{k=1}^\infty \frac{\Omega
^{2}}{ \omega _k^2-\Omega ^2}. \label{Nelson3}
\end{equation}
We see that, in this limit, the above procedure is exactly the
analogue of mass renormalization in quantum field theory: the
addition of a counterterm $- N \eta^2 q_0^2$ allows one to
compensate the infinity of $\omega _0^2$ in such a way as to leave a
finite, physically meaningful, renormalized frequency
$\bar{\omega}$.

To proceed, we take the constant $\eta $ as
\begin{equation}
\eta =\sqrt{\frac{4g\Delta \omega}{\pi} }, \label{eta}
\end{equation}
where $\Delta \omega $ is the interval between two neighboring field
frequencies and $g$ is the coupling constant with dimension of
frequency. The environment frequencies $\omega _k$ can be written in
the form
\begin{equation}
\omega _k=k\frac{\pi c}R,\;\;\;\;k=1,2,\ldots \;,  \label{discreto}
\end{equation}
and, so, $\Delta \omega = \pi c/R$. Then, using the identity
\begin{equation}
\sum_{k=1}^\infty \frac 1{k^2-u^2}=\frac 12\left[ \frac 1{u^2}-\frac
\pi u\cot \left( \pi u\right) \right] ,  \label{id4}
\end{equation}
Eq.~(\ref{Nelson3}) can be written in closed form:
\begin{equation}
\cot \left( \frac{R\Omega }c\right) =\frac \Omega {2 g}+\frac
c{R\Omega }\left( 1-\frac{R\bar{\omega}^2}{2 gc}\right).
\label{eigenfrequencies1}
\end{equation}
The elements of the transformation matrix, turning the particle--field
system to principal axis, are obtained in terms of the physically
meaningful quantities $\Omega _r$, $\bar{\omega}$, after some rather
long but straightforward manipulations \cite{adolfo1}. They read
\begin{eqnarray}
t_0^r &=&\frac{\eta \Omega _r}{\sqrt{(\Omega
_r^2-\bar{\omega}^2)^2+\frac{ \eta ^2}2(3\Omega
_r^2-\bar{\omega}^2)+4g^2\Omega _r^2}},
\label{t0r2} \\
t_k^r &=&\frac{\eta \omega _k}{\omega _k^2-\Omega _r^2}t_0^r.
\label{t0r21}
\end{eqnarray}

Let us now consider the eigenstates of our system, $\left|
n_0,n_1,n_2,...\right\rangle $, represented by the normalized
eigenfunctions, written in terms of the normal coordinates
$\{Q_r\}$,
\begin{equation}
\phi _{l_0l_1l_2...}(Q,t) = \prod_s\left[
\sqrt{\frac{2^{l_s}}{l_s!}} H_{l_s}\left( \sqrt{\frac{\Omega
_s}\hbar }Q_s\right) \right] \Gamma _0 \,
e^{-i\sum_s\left(l_s+\frac{1}{2}\right)\Omega _st},
\label{autofuncoes}
\end{equation}
where $H_{l_s}$ stands for the $l_s$-th Hermite polynomial and
$\Gamma _0$ is the normalized vacuum eigenfunction.

We introduce \textit{renormalized} coordinates $
q_0^{\prime }$ and $\{q_i^{\prime }\}$ for the \textit{dressed }
particle and the \textit{dressed} field, respectively, defined by
\begin{equation}
\sqrt{\bar{\omega}_\mu }q_\mu ^{\prime }=\sum_rt_\mu ^r \sqrt{\Omega
_r}Q_r, \label{qvestidas1}
\end{equation}
where $\bar{\omega}_\mu =\{\bar{\omega},\;\omega _i\}$. In terms of
the renormalized coordinates, we define for a fixed instant, $t=0$,
\textit {dressed} states, $\left| \kappa _0,\kappa _1,\kappa _2,...
\right\rangle $ by means of the complete orthonormal set of
functions \cite{adolfo1}
\begin{equation}
\psi _{\kappa _0\kappa _1...}(q^{\prime })=\prod_\mu \left[
\sqrt{\frac{ 2^{\kappa _\mu }}{\kappa _\mu !}}H_{\kappa _\mu }\left(
\sqrt{\frac{\bar{ \omega}_\mu }\hbar }q_\mu ^{\prime }\right)
\right] \Gamma _0, \label{ortovestidas1}
\end{equation}
where $q_\mu ^{\prime }=\left\{ q_0^{\prime },\,q_i^{\prime
}\right\} $. Note that the ground state $\Gamma _0$ in the above
equation is the same as in Eq.~(\ref{autofuncoes}). The invariance
of the ground state is due to our definition of renormalized
coordinates given by Eq.~(\ref{qvestidas1}). Each function $\psi
_{\kappa _0\kappa _1...}(q^{\prime })$ describes a state in which
the dressed oscillator $q_\mu ^{\prime }$ is in its $\kappa _\mu
$-th excited state. In terms of the bare coordinates, the
renormalized coordinates are expressed as
\begin{equation}
q_\mu ^{\prime }=\sum_\nu \alpha _{\mu \nu }q_\nu ,
\label{qvestidas3}
\end{equation}
where
\begin{equation}
\alpha _{\mu \nu }=\frac 1{\sqrt{\bar{\omega}_\mu }}\sum_rt_\mu
^rt_\nu ^r \sqrt{\Omega _r}.  
\label{qvestidas4}
\end{equation}

Remark that the introduction of the renormalized coordinates
implies, differently from the bare vacuum, the stability of the
dressed vacuum state since, by construction, it is identical to the
ground state of the interacting Hamiltonian (\ref{diagonal}). 
Also it is important to notice that the renormalized coordinates and the
dressed states, defined in Eqs.~(\ref{qvestidas1}) and
(\ref{ortovestidas1}) are \textit{new}
collective objects, \textit{different} from the normal coordinates
$Q$ and the eigenstates (\ref{autofuncoes}). Since the transformation 
(\ref{qvestidas1}) is not orthogonal, the Hamiltonian is not diagonal in the 
renormalized coordinates. Thus, distinctly from the
eigenstates, our dressed states are all unstable, except for the
ground dressed ($\{\kappa _\mu =0\}$) state. We shall assume that the
dressed states (\ref{ortovestidas1}) are the physically appropriate states  
to describe the time evolution of superpositions of states of the mechanical 
oscillator, taking into account non-perturbatively the effect of the interaction
with the environment. This is an alternative to the use of states written in 
terms of the bare coordinates $q_\mu $, which would require a 
perturbative renormalization procedure to correct order by order 
the oscillator frequency.

Let us consider, for instance, the particular dressed state $\left| \Gamma
_{1}^{\mu}(0)\right\rangle $, represented by the wave function $\psi
_{00\cdots 1(\mu )0\cdots }(q^{\prime })$. It can be seen as describing, at a given instant, the
configuration in which {\it only} the $\mu$-th dressed oscillator
is in the ``first excited level", all others being in their
``ground states". These ``levels" should not be confused with the {\it stationary} states given by Eq.~(\ref{autofuncoes}); from now on, we shall use such a terminology to facilitate the reference to dressed states. 
 As shown in Ref.~\cite{adolfo1}, the time evolution
of the state $\left| \Gamma _{1}^{\mu} \right\rangle$ is given by
\begin{equation}
\left|\Gamma _{1}^{\mu} (t)\right\rangle = \sum_\nu f_{\mu \nu
}(t)\left| \Gamma _{1}^{\nu} (0)\right\rangle ,
\label{ortovestidas51}
\end{equation}
where
\begin{equation}
 f_{\mu \nu }(t) = \sum_st_\mu ^st_\nu ^se^{-i\Omega _st} .
\label{ortovestidas5}
\end{equation}
Moreover, it can be shown that, for all $\mu$,
\begin{equation}
\sum _{\nu}\left|f_{\mu \nu}(t)\right|^{2}=1 , \label{probabilidade}
\end{equation}
which allows to interpret the coefficients $f_{\mu \nu }(t)$ as
probability amplitudes; for example, $f_{0 0}(t)$ is the probability
amplitude that, if the dressed particle is in the first excited
state at $t=0$, it remains excited at time $t$, while $f_{0 i}(t)$
represents the probability amplitude that the $i$-th dressed
harmonic mode of the cavity be at the first excited level.  Also,
the elements of the matrix
$T_{\kappa_0\kappa_1\cdots}^{l_0l_1\cdots}$, connecting the dressed
states to the eigenstates, can be evaluated. For the dressed state in
which only the $\kappa_{\mu}$-th dressed oscillator is in the $N$-th
excited level, all other being in the ground state, we have
\cite{adolfo1},
\begin{equation}
T_{0,0,\cdots}^{l_0l_1\cdots}=
\left(\frac{N!}{l_{0}!\,l_{1}!\cdots}\right)^{\frac{1}{2}}
\left(t_{\mu}^{0}\right)^{l_0}\left(t_{\mu}^{1}\right)^{l_1}\cdots,
\label{T}
\end{equation}
where $l_0+l_1+\cdots =N$.

\section{Behavior of the superposition of states in a cavity}

We now analyze the situation in which, at time $t=0$, the
environment oscillators are in their ground states and the particle
is in a superposition of dressed states, and the whole system is
allowed to evolve in time. Specifically, we take, as the initial
particle dressed state, an arbitrary superposition of the
first-excited and the ground states,
\begin{equation}
\left| \psi \right\rangle = \sqrt{\xi} \left| \Gamma _{1}^{0}
(0)\right\rangle + \sqrt{1 - \xi} e^{i\phi} \left| \Gamma
_{0}\right\rangle , \label{superposicao1}
\end{equation}
where $0 < \xi < 1$. While the ground state is stable, the particle
first-excited state evolves in time according to
Eq.~(\ref{ortovestidas51}), that is, $\left|\Gamma _{1}^{0}
(t)\right\rangle = \sum_\nu f_{0 \nu }(t)\left| \Gamma _{1}^{\nu}
(0)\right\rangle$ with $f_{0 \nu }(t) = \sum_st_0 ^st_\nu
^se^{-i\Omega _st}$. At the instant $t$, the state of the system is
given by the density matrix
\begin{equation}
\varrho(t)=e^{-iHt}|\psi\rangle\langle\psi|e^{iHt} . \label{ga1}
\end{equation}

We are interested in studying the influence of the environment on
the time evolution of the state of the subsystem corresponding to
the particle. Thus, we take the trace over all the degrees of
freedom associated with the field, that is, we consider the reduced
density matrix
\begin{equation}
\rho(t)=\sum_{\{ k_i=0 \} }^\infty \langle k_1, k_2,...|
e^{-iHt}|\psi\rangle \langle\psi|e^{iHt}| k_1, k_2,...\rangle ,
\label{ga2}
\end{equation}
whose elements between dressed particle states are
\begin{equation}
\rho_{mn}(t)=\langle m|\rho|n\rangle =\sum_{\{  k_i=0 \} }^\infty
\langle m, k_1, k_2,...| e^{-iHt}|\psi\rangle \langle\psi|e^{iHt}|n,
k_1, k_2,...\rangle. \label{ga3}
\end{equation}
Replacing Eq.~(\ref{superposicao1}) in Eq.~(\ref{ga3}) we get
\begin{eqnarray}
\rho_{mn}(t)&=& \sum_{\{  k_i=0 \}}^\infty\left\{ \xi{\mathcal
A}_{100...}^{m k_1 k_2...}(t) {\mathcal A}_{100...}^{n k_1
k_2...\ast}(t)  + \, (1-\xi) {\mathcal A}_{000...}^{m k_1
k_2...}(t){\mathcal
A}_{000...}^{n k_1 k_2...\ast}(t) \right. \nonumber\\
 & & + \, \sqrt{\xi (1-\xi)}  e^{-i\phi} {\mathcal
A}_{100...}^{m k_1 k_2...}(t) {\mathcal A}_{000...}^{n k_1
k_2...\ast}(t) \nonumber \\
 & &  \left.  +\, \sqrt{\xi (1-\xi)}
e^{i\phi} {\mathcal A}_{000...}^{m k_1 k_2...}(t){\mathcal
A}_{100...}^{n k_1 k_2...\ast}(t)  \right\},  \label{ga4}
\end{eqnarray}
where
\begin{equation}
{\mathcal A}_{n_0n_1n_2...}^{m_0m_1m_2...}(t) = \langle
m_0,m_1,m_2,... |e^{-iHt} |n_0,n_1,n_2,...\rangle \label{ga5}
\end{equation}
are the probability amplitudes between the dressed states. These
probability amplitudes can be computed using
Eqs.~(\ref{ortovestidas5}) and (\ref{T}). We then have
\begin{equation}
{\mathcal A}_{000...}^{m_0m_1m_2...}(t)=e^{-iE_0t}
\delta_{0m_0}\delta_{0m_1}\delta_{0m_2}..., \label{ga6}
\end{equation}
\begin{equation}
{\mathcal
A}_{100...}^{m_0m_1m_2...}(t)=e^{-iE_0t}\sum_{\mu}f_{0\mu}(t)
\delta_{1m_\mu} \prod_{\nu\neq\mu}\delta_{0m_\nu}, \label{ga7}
\end{equation}
where $E_0=\sum_r\Omega_r/2$ is the ground state energy of the
system. Substituting the above equations into Eq.~(\ref{ga4}), we
obtain, after summing over the $ k_i$'s,
\begin{eqnarray}
\rho_{mn}(t) & = & \left( 1 - \xi + \xi \sum_k |f_{0k}(t)|^2 \right)
\delta_{0m}\delta_{0n}   + \,
\xi |f_{00}(t)|^2 \delta_{1m} \delta_{1n}  \nonumber \\
 & & + \, \sqrt{\xi(1-\xi)} e^{-i\phi}
 f_{00}(t)\delta_{1m}\delta_{0n}  + \, \sqrt{\xi(1-\xi)}
e^{i\phi} f_{00}^\ast(t) \delta_{0m}\delta_{1n}. \\ \label{ga8}
\end{eqnarray}
Using that $\sum_\mu|f_{0\mu}|^2=1$, the nonzero elements of the
reduced density matrix are given by
\begin{eqnarray}
\rho_{00}(t)&=&1-\xi|f_{00}(t)|^2,\label{ga91}\\
\rho_{11}(t)&=&\xi|f_{00}(t)|^2,\label{ga92}\\
\rho_{10}(t)&=&\sqrt{\xi(1-\xi)} e^{-i\phi} f_{00}(t),\label{ga93}\\
\rho_{01}(t)&=&\sqrt{\xi(1-\xi)} e^{i\phi}f_{00}^\ast(t).
\label{ga94}
\end{eqnarray}
Note that ${\rm Tr}[\rho(t)] = 1$, which means that the reduced
density matrix $\rho(t)$ does represent a physical state of the
particle; also, it is not a pure state, since ${\rm Tr}[\rho^2(t)]
\neq 1$. The degree of impurity of a quantum state can be quantified
by the departure of the idempotent property; in the present case, we
have
\begin{equation}
D(t;\xi) = 1 - {\rm Tr}[\rho^2] = 2 \xi^2 |f_{00}(t)|^2 \left( 1 -
|f_{00}(t)|^2 \right) . \label{DI}
\end{equation}
We are thus left with the calculation of the quantity $f_{00}(t)$,
which will be done for the particular situations of an arbitrarily
large cavity ($R\rightarrow\infty$) and of a small cavity.

\subsection{Large cavity}

For an arbitrarily large value of $R$, Eq.~(\ref{t0r2}) reduces to
\begin{equation}
t_0^r\rightarrow \lim_{R\rightarrow \infty
}\frac{\sqrt{4g/\pi}\,\Omega\, \sqrt{\pi c/R}}{\sqrt{(\Omega
^2-\bar{\omega}^2)^2+4g^2\Omega ^2}}. \label{t0r(R)}
\end{equation}
Also, in this limit, $\Delta \omega = \pi c/R \rightarrow d\omega
=d\Omega $ and the expression for the quantity $f_{00}(t)$, given by
Eq.~(\ref{ortovestidas5}), can be cast in the form
\begin{equation}
f_{00}(t)=\frac{4g}{\pi}\int_0^\infty d\Omega \, \frac{\Omega
^2e^{-i\Omega t} }{(\Omega ^2-\bar{\omega}^2)^2+4g^2\Omega ^2}.
\label{f00}
\end{equation}

\begin{figure}[ht]
\begin{center}
\scalebox{0.78}{{\includegraphics{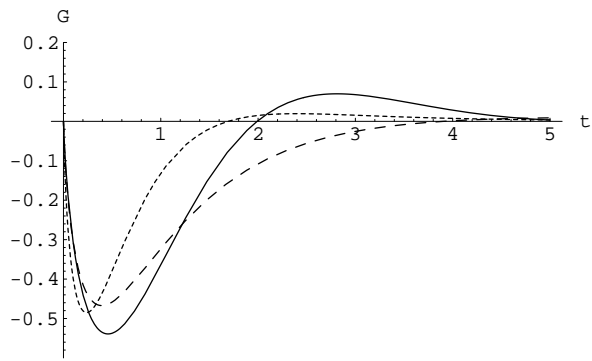}}}
\end{center}
\caption{Behavior of $G(t;\bar{\omega},g)$ as a function of $t$ for
the three distinct cases: $\bar{\omega} = 1.5$ and $g=1.0$ (full
line); $\bar{\omega} = 1.0$ and $g=1.2$ (dashed line); and
$\bar{\omega} = g = 2.0$ (dotted line), in arbitrary units.}
\label{G}
\end{figure}

\begin{figure}[ht]
\begin{center}
\scalebox{0.80}{{\includegraphics{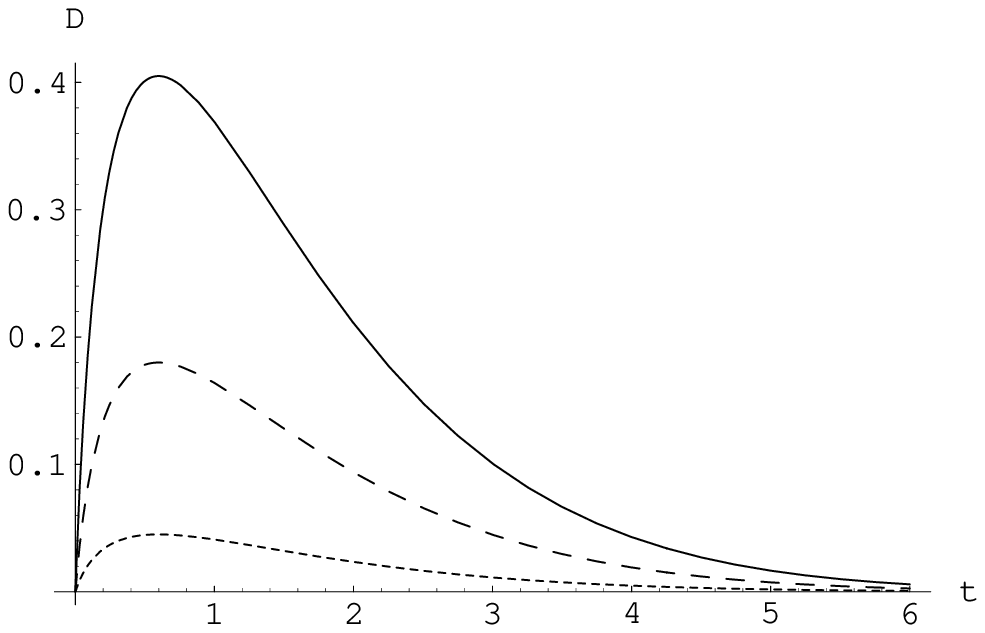}}}
\end{center}
\caption{Behavior of $D(t;\xi)$ as a function of $t$, with
$\bar{\omega} = 1.0$ and $g = 0.5$ fixed (in arbitrary units), for
some values of $\xi$: $0.3$, $0.6$ and $0.9$ (dotted, dashed and
full lines, respectively).} \label{D}
\end{figure}

The real part of $f_{00}(t)$ can be evaluated directly using the
residue theorem. Defining the parameter $\kappa$ by
\begin{equation}
\kappa =\sqrt{\bar{\omega}^2-g^2} ,  \label{k}
\end{equation}
we obtain, for the three distinct cases,

\noindent a) $\kappa ^2>0:$
\begin{equation}
f_{00}(t) = e^{-gt}\left[ \cos{\kappa t} - \frac{g}{\kappa}
\sin{\kappa t} \right] + i G(t;\bar{\omega},g) ; \label{eq27}
\end{equation}

\noindent b) $\kappa ^2=0:$
\begin{equation}
f_{00}(t)= e^{-gt} \left[ 1 - g t \right] + i G(t;\bar{\omega},g);
\label{eq28}
\end{equation}

\noindent and c) $\kappa ^2<0:$
\begin{equation}
f_{00}(t) = e^{- g t} \left[ \cosh{|\kappa|t}
 - \frac{g}{|\kappa|} \sinh{|\kappa|t} \right] + i
G(t;\bar{\omega},g) , \label{eq29}
\end{equation}
where the function $G(t;\bar{\omega},g)$ is given by
\begin{equation}
G(t;\bar{\omega},g) = - \frac{4g}{\pi}\int_0^\infty
dy\frac{y^2\sin{yt}}{(y^2 - \bar{\omega}^2)^2 + 4 g^2 y^2} \;.
\label{J}
\end{equation}
The overall behavior of the function $G(t;\bar{\omega},g)$ is
illustrated in Fig.~\ref{G}; we see then that, in all cases, the
real and imaginary parts of $f_{00}(t)$ decay with the time, for
large times. This aspect dictates the behavior of the degree of
purity, as a function of time, as illustrated in Fig.~\ref{D}, for a
case with $g<\bar{\omega}$; similar results are obtained for the
other regimes.

Replacing $f_{00}(t)$, given by (\ref{eq27}), (\ref{eq28}) or
(\ref{eq29}), into Eqs.~(\ref{ga91})--(\ref{ga94}) leads to the
elements of the reduced density matrix; simpler expressions are
obtained for large times ($t\gg 1/\bar{\omega}$) at both
\textit{weak}- and \textit{strong-coupling} regimes. For large $t$,
$G(t;\bar{\omega},g)$ can be approximated as
\begin{equation}
G(t;\bar{\omega},g)\approx
\frac{8g}{\pi\bar{\omega}^4t^3}\;\;\;\;(t\gg \frac 1{\bar{\omega}
}).  \label{J1}
\end{equation}
Thus, in the limit case of weak coupling between the particle and
the environment, $g\ll\bar{\omega}$ (corresponding to
$\kappa\approx\bar{\omega}$), the large-time approximation of the
reduced density matrix gives
\begin{eqnarray}
\rho_{11}(t) & \approx & \xi \left \{ e^{-2gt}  \left[ \cos
\bar{\omega} t - \frac{g}{\bar{\omega}}
 \sin \bar{\omega} t \right]^2 + \frac{64g^2}{\pi^2 \bar{\omega}^{8}
 t^6} \right\} ,  \label{densidade31}  \\
\rho_{00}(t) & = & 1-\rho_{11}(t) , \label{densidade3} \\
\rho _{10}(t) & \approx & \sqrt{\xi(1-\xi)} e^{-i\phi} \left \{
e^{-gt} \left[ \cos \bar{\omega} t - \frac{g}{\bar{\omega}}
 \sin \bar{\omega} t \right] + i\frac{8g}{\pi\bar{\omega}^4t^3}
 \right\} , \label{fraca1} \\
\rho_{01}(t) & = & \rho_{10}^{*}(t) . \label{fraca2}
\end{eqnarray}

Similar results can be obtained for the case of a strong coupling
between the particle and the environment, that is, when
$g\gg\bar{\omega}$ (i.e. $|\kappa|\approx g$); for large $t$, we
find
\begin{eqnarray}
\rho _{11}(t) & \approx & \xi \left[ e^{-4 g t}  +
\frac{64g^2}{\pi^2\bar{\omega}^8t^6} \right] ,
 \label{densidade41} \\
\rho _{00}(t) & = & 1-\rho_{11}(t) , \label{densidade4} \\
\rho _{10}(t) & \approx & \sqrt{\xi(1-\xi)} e^{-i\phi}  \left[ e^{-2
g t} + i\frac{8g}{\pi\bar{\omega}^4t^3}\right] ,
\label{rhored10} \\
\rho _{01}(t) & = & \rho_{10}^{*}(t) . \label{rhored01}
\end{eqnarray}

We see from the above equations that, for both weak- and
strong-coupling regimes, in the case of a very large cavity,  except
for the vacuum--vacuum component which goes to $1$, all other matrix
elements of the reduced density matrix tend to zero as $t\rightarrow
\infty $. This means that the particle mixed state $\rho(t)$ tends
to the dressed-particle vacuum state as time evolves; thus, for
large cavities, the particle system is dissipative. This is also
illustrated in Fig.~\ref{D}, with $D(t;\xi)$ vanishing as
$t\rightarrow\infty$ for all values of $\xi$. Notice that the the
real and imaginary parts of the matrix elements $\rho_{10}(t)$ and
$\rho_{01}(t)$, which corresponds to interference terms, decay more
slowly then the element $\rho_{11}(t)$.

\subsection{Small cavity}

For a finite cavity, the spectrum of eigenfrequencies is discrete,
$\Delta\omega$ is large, and so the approximation made in the case
of large cavity does not apply; no analytical result can be obtained
for $f_{00}(t)$, given by Eq.~(\ref{ortovestidas5}), in this case.
However, if the cavity is sufficiently small, the frequencies
$\Omega_r$ can be determined as follows \cite{adolfo2}. In terms of
the small dimensionless parameter
\begin{equation}
\delta = \frac{g}{\Delta\omega} = \frac{gR}{\pi c} , \label{delta}
\end{equation}
Eq.~(\ref{eigenfrequencies1}) is rewritten as
\begin{equation}
\cot(\pi\theta) = \frac{\theta}{2\delta} + \frac{1}{\pi\theta}
\left( 1 - \frac{\pi\bar{\omega}^2 \delta}{2 g^2} \right),
\label{eignfreqSC}
\end{equation}
where $\theta = \Omega /\Delta\omega$. With $\delta\ll 1$, which
corresponds to $R\ll\pi c/g$ (a small cavity), as plots of both
sides of Eq.~(\ref{eignfreqSC}) as functions of $\theta$ are drawn,
one sees that, for $k=1,2,\ldots$, the solutions of
Eq.~(\ref{eignfreqSC}) can be written as $\theta_k \approx k +
\epsilon_k$ with $\epsilon_k \ll 1$; then, expanding $\cot(\pi
\epsilon_k)$ for small $\epsilon_k$, one finds
\begin{equation}
\Omega_k \approx \Delta \omega \left( k + \frac{2\delta}{\pi k}
\right) = \frac{g}{\delta} \left( k + \frac{2\delta}{\pi k} \right).
\label{OmegaK}
\end{equation}
If we further impose that $\delta < 2g^2/\pi\bar{\omega}^2$, a
condition compatible with $\delta\ll 1$, then $\Omega_0$ is found to
be very close to $\bar{\omega}$, that is,
\begin{equation}
\Omega_0 \approx \bar{\omega} \left( 1 - \frac{\pi \delta}{2}
\right) . \label{Omega0}
\end{equation}

To determine $f_{00}(t)$, we have to calculate the square of the
matrix elements $(t_0^0)^2$ and $(t_k^0)^2$. From Eq.~(\ref{t0r21}),
using that $\omega_k = k \Delta \omega = kg/\delta$, we find (to
first order in $\delta$)
\begin{equation}
(t_k^0)^2 \approx \frac{4\delta}{\pi k^2}\; (t_0^0)^2 . \label{tk02}
\end{equation}
Now, considering the orthonormalization condition (\ref{normcond}),
we determine $(t_0^0)^2$ as
\begin{equation}
(t_0^0)^2 \approx \left( 1 + \frac{4 \delta}{\pi}
\sum_{k=1}^{\infty} k^{-2} \right)^{-1} = \left( 1 +
\frac{2}{3}\pi\delta \right)^{-1} ,
\end{equation}
where we have used that $\zeta(2) = \sum_{k=1}^{\infty} k^{-2} =
\pi^2/6$. We thus obtain, for sufficiently small cavities ($\delta
\ll 1$),
\begin{equation}
f_{00}(t) \approx  \left( 1 + \frac{2}{3}\pi\delta \right)^{-1}
\left[ e^{-i \bar{\omega} \left( 1 - \frac{\pi\delta}{2}
 \right)t}
 + \sum_{k=1}^\infty \frac{4\delta}{\pi k^2}
e^{-i\frac{g}{\delta}\left(k + \frac{2\delta}{\pi k} \right) t}
\right] .  \label{f00peqN}
\end{equation}
Notice that, if we had calculated $(t_0^0)^2$ by approximating
directly Eq.~(\ref{t0r2}) for small $\delta$, instead of using the
normalization condition (\ref{normcond}), $f_{00}(t)$ would not
satisfy the requirement $\left| f_{00}(t)\right|^2\leq 1$ with the
equality occurring when $t=0$. Also, it is worth mentioning that the
approximation we made holds independently of the strength of the
coupling between the particle and the environment and, thus, it
applies for both weak and strong limits.

Substituting $f_{00}(t)$, given by Eq.~(\ref{f00peqN}), into
Eqs.~(\ref{ga91})--(\ref{ga94}) leads to the elements of the reduced
density matrix for the case of a small cavity. Distinctly from the
case of a very large cavity, now, these matrix elements do not tend
to zero as time evolves; in fact, they oscillate, never reaching
zero. In particular, we find that
\begin{eqnarray}
\rho_{11}(t) & \approx & \xi \left( 1 + \frac{2}{3}\pi\delta
\right)^{-2} \left\{ 1 + \frac{8\delta}{\pi} \sum_{k=1}^{\infty}
\frac{1}{k^2} \cos{
 \left[ \bar{\omega} (1 - \frac{\pi\delta}{2}) - \frac{g}{\delta}
 (k + \frac{2\delta}{\pi k}) \right] t }  \right. \nonumber \\
  & & + \left. \frac{16 \delta^2}{\pi^2} \sum_{k,l=1}^{\infty}
  \frac{1}{k^2 l^2} \cos{  \left[ ( \frac{g}{\delta} -
  \frac{2 g}{\pi k l}) (k-l) \right] t } \right\} .
   \label{rho11}
\end{eqnarray}
Note that $\rho_{11}(t)$ is an oscillating, nonperiodic, function
of time. A lower bound for $\rho_{11}$ can be found if one replaces
the cosine terms in Eq.~(\ref{rho11}) by $-1$; in fact, to order
$\delta^2$, one finds
\begin{equation}
\rho_{11}(t) > \xi \left[ 1 - \frac{8}{3}\, \pi \delta +
\frac{8}{9}\, \pi^2 \delta^2 \right] . \label{lowerbound}
\end{equation}
This means that the particle state never reduces to the
dressed-particle vacuum state; the particle system, maintaining
itself in a mixed state, keeps exchanging quanta with the
environment.

\begin{figure}[ht]
\begin{center}
\scalebox{0.80}{{\includegraphics{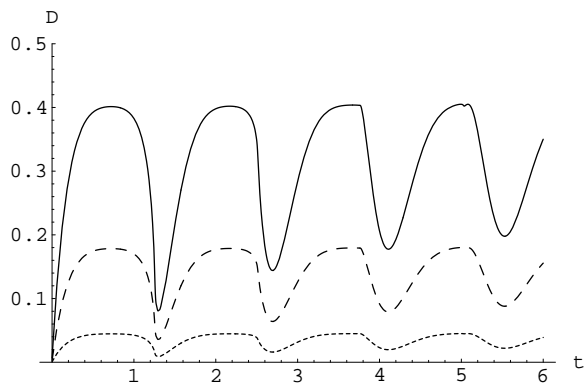}}}
\end{center}
\caption{Behavior of $D(t;\xi)$ as a function of $t$ for $\delta =
0.1$, with $\bar{\omega} = 1.0$ and $g = 0.5$ fixed (in arbitrary
units), for some values of $\xi$: $0.3$, $0.6$ and $0.9$ (dotted,
dashed and full lines, respectively).} \label{DSC}
\end{figure}

The degree of impurity of this particle state at time $t$ can be
calculated using Eq.~(\ref{DI}) or, equivalently, with
\begin{equation}
D(t;\xi) = 2\, \rho_{11}(t) \left[ \xi - \rho_{11}(t) \right] .
\label{DI2}
\end{equation}
This quantity is plotted in Fig.~\ref{DSC}, for the same values of
$\bar{\omega}$, $g$ and $\xi$ as those used in Fig.~\ref{D} (for
comparison), fixing a specific value of $\delta$; one clearly sees
the oscillatory, nonperiodic, behavior mentioned above. Therefore,
in the case of a small cavity, for both weak- and strong-coupling
regimes, the particle system is not dissipative; it keeps exchanging
energy with the environment. For example, in the case of weak
coupling, a physically interesting situation occurs when
interactions of electromagnetic type are involved. In this case, we
can take $g=\alpha \bar{\omega}$, where $\alpha$ is the
fine-structure constant, $\alpha \sim 1/137$. For a frequency
$\bar{\omega}\sim 2\times 10^{11}$/s (in the microwave band) and a
cavity with $R\sim 10^{-2}$ m, we have $\delta \sim 0.016$ and, so,
the diagonal element $\rho _{11}(t)$ will never fall below the value
$\sim 0.87\xi$. For strong coupling, $g=\beta\, \bar{\omega}$
($\beta >1$), a similar result holds for a smaller cavity. The same
kind of oscillatory behavior can be verified for the nondiagonal
elements of the density matrix. In other words, the system confined
in a small cavity is not dissipative, contrary to the case of a
large cavity.

\section{Concluding Remarks}

We have presented in this paper a nonperturbative treatment of a
quantum system consisting of a particle, in the harmonic
approximation, coupled to an environment modeled by noninteracting
oscillators. We have used renormalized coordinates in terms of which
dressed states can be constructed. These states allow to naturally
separate the system into the dressed (physically observed) particle
and the dressed environment by means of the conveniently chosen
renormalized coordinates, $q_0'$ and $q_j'$, associated respectively
to the dressed particle and to the dressed oscillators composing the
environment. The dressed particle will contain automatically all the
effects of the environment in it. Using this formalism we perform a
nonperturbative study of the time evolution of a superposition of
the ground state and the first excited state of the particle, the
system being confined in a perfectly reflecting cavity of radius
$R$.

In the $R\rightarrow\infty$ limit, we find dissipation (damping)
with dominance of the interference terms of the density matrix, in
both weak- and strong-coupling regimes. For small values of $R$, the
system is not dissipative, presenting stability. In both cases we
have obtained results in agreement with expected behaviours.
However, we do not simply recover well-known features with our
method. For instance, as far as we know, the corrections to the
exponential decay in Eqs.~(\ref{densidade31}) to (\ref{fraca2}) and
from Eqs.~(\ref{densidade41}) to (\ref{rhored01}) have not been
reported before in the literature. Also, the absence of dissipation
in a small cavity can be understood as the analogue of spontaneous
emission inhibition for an atom, which has a qualitative explanation
when the lowest resonance cavity frequency is larger than the
oscillator frequency. But also in this case we are able, with our
method, to give rigorous expressions for the elements of the density
matrix (in an analogous manner as has been done in \cite{adolfo2}
for the size of cavities ensuring stability of excited atoms).

It is worth to stress that the renormalized coordinates and the dressed states, defined in Eqs.~(\ref{qvestidas1}) and
(\ref{ortovestidas1}), allow an exact treatment of the problem, in
which one completely avoids the use of perturbation theory. As already noticed in the text below Eq.~(\ref{qvestidas4}), the
renormalized coordinates and dressed states are \textit{new}
collective objects, \textit{different} from the normal coordinates
$Q$ and the eigenstates (\ref{autofuncoes}). Distinctly from the
eigenstates, our dressed states are all unstable, except for the
ground dressed ($\{\kappa _\mu =0\}$) state. We assume that the
dressed states (\ref{ortovestidas1}) are the physically meaningful
states, instead of the ones written in terms of the bare coordinates
$q_\mu $. This can be seen as analogous to the wave-function
renormalization in quantum field theory, which justifies referring
to $q^{\prime}_{\mu}$ as \textit{renormalized} coordinates. Also, it
is worth mentioning that the introduction of renormalized
coordinates naturally ensures that the dressed vacuum state is
stable, contrarily to the bare perturbative vacuum, since, by
construction, it is identical to the ground state of the interacting
Hamiltonian (\ref{diagonal}). Remember that the invariance of the
ground state is due to our definition of renormalized coordinates
given by Eq.~(\ref{qvestidas1}).

As already explained in the Introduction, our formalism and results
are restricted to a linear model, but it has the advantage of being
nonperturbatively solvable. More realistic physical situations,
however, would require a nonlinear coupling between the particle and
the environment. It is nevertheless envisageable that such a study
of complex nonlinear systems could be handled with our
nonperturbative method. Indeed, the concept of renormalized
coordinates has already  been extended to a nonlinear model in
Ref.~\cite{ga1}. Also, the generalization of the work presented in
the present paper to the case of finite temperature, as well as the
study of dressed coherent states, is in progress and will be
presented elsewhere.

\ack

This work was partially supported by CNPq/MCT, Brazil. GFH thanks 
FAPEMIG/MG, Brazil, for financial support. APCM thanks FAPERJ for partial support.\\

\end{document}